\documentclass[aps,prb,twocolumn,groupedaddress,showpacs]{revtex4}
\usepackage{epsfig}
\usepackage[dvipsnames,usenames]{color}
\usepackage{color}
\usepackage{amsmath}

\emergencystretch=\maxdimen
\begin{document}
\title{Effects of Strong Interactions in a Half Metallic Magnet:\\
a Determinant Quantum Monte Carlo Study}
\author{M.~Jiang$^{1,2}$, W.E.~Pickett,$^{1}$ and R.T.~Scalettar$^1$}

\affiliation{$^1$Physics Department, University of California, Davis,
  California 95616, USA}
\affiliation{$^2$Department of Mathematics, University of California, Davis,
  California 95616, USA}

\begin{abstract}
Understanding the effects of electron-electron interactions in half metallic
magnets (HMs), which have band structures with one gapped spin channel
and one metallic channel,
poses fundamental theoretical issues as well as having
importance for their potential applications.  Here we use determinant
quantum Monte Carlo to study the impacts of an on-site Hubbard
interaction $U$, finite temperature, and an external (Zeeman) magnetic
field on a bilayer tight-binding model which is a half-metal in the absence of
interactions,  by calculating the spectral density, conductivity, spin
polarization of carriers, and local magnetic properties.  We quantify
the effect of $U$ on the degree of thermal depolarization, and follow
relative band shifts and monitor when significant gap states appear,
each of which can degrade the HM character.  For this model,
Zeeman coupling induces, at fixed particle number, two successive
transitions: compensated half-metal with spin-down band
gap $\rightarrow$ metallic ferromagnet $\rightarrow$ saturated ferromagnetic insulator.
However, over much of the more relevant parameter regime, the
half-metallic properties are rather robust to $U$.
\end{abstract}
\pacs{71.10.Fd, 71.30.+h, 02.70.Uu}
\maketitle

\section{Introduction and Background}

As a family of promising candidates for spintronics application,
half-metallic (HM) magnetic materials have attracted much interest over
the past two decades\cite{review,pickettreview}.  de Groot {\it et
al.}\cite{halfmetal} discovered HMs computationally and popularized the
HM possibility especially in Heusler and half-Heusler compounds.
The distinguishing characteristic of a HM is that at the mean field level the Fermi level for one spin direction falls within a gap
for the other direction, resulting in 100\% polarized conduction and
obvious spintronics applications.
Motivated by the unusual magneto-optical properties, the Heusler class of
intermetallic materials provided the most promising realizations of HM
character.  Thermal fluctuations degrade
spin alignment and thus destroy the ideal HM.  In addition, real
materials effects such as crystal imperfections can couple spins and
degrade the polarization; spin-orbit coupling destroys the true HM.
These phenomena can be minor at low temperatures compared with the Curie
temperature (which ranges 500-730 K in the half Heusler family NiMnZ
(Z=Co, Pd, Pt, Sb)~[\onlinecite{500K}] and can be higher), but because
proposed applications do not rely on 100\% polarization, HMs remain
viable for near-future electronic devices.

A variety of experimental techniques, including positron annihilation,
spin-resolved photoemission spectroscopy, Andreev reflection, and nuclear
magnetic resonance, have been employed to assess the
character and polarization level of proposed HMs, often with less than
definitive results.  Comparision is made with density functional theory
(DFT) based electronic structure
calculations, which still play the dominant role in the specification of
HMs and in the search for additional HMs and for the even more exotic
half metallic antiferromagnets\cite{HMAF,WEPhmaf,HMAF_Xu}.  HMAFs, better
characterized as compensated HMs, have zero macroscopic moment so they
can provide additional functionalities.
Known or strongly anticipated HMs now span a diverse collection of
materials with different chemical and physical properties.

An important question about HMs, not yet well clarified, is the impact
of dynamic interactions (which lie beyond
DFT methods) on the character
and the survival of the single-spin-channel gap that defines the HM
phase.  One of the most fundamental consequences of repulsive on-site interactions
is local moment formation and dynamics, the study of whose effects in metals has a
long history (Anderson impurities, Kondo systems). Local moments in
insulators are also rather well studied and present a largely distinct
set of issues (gap states versus band resonances, magnetic
activity)\cite{fazekas99}.  HMs, especially in oxides,
bring in all of these issues: although one spin is gapped,
it is not electrically an insulator since there is metallic screening,
but the two spin channels are
fundamentally distinct. There have now been a number of studies, in
particular by Katsnelson, Chioncel, and collaborators, that indicate
degradation of the HM gap -- sometimes severely -- by electron-electron
interactions.  One general picture is that interactions, typically
pictured in terms of a magnetic polaron formed by charge
carrier-magnon binding, lead to tailing of spectral density into the
gap and sometimes to mid-gap states that would substantially degrade
performance as a HM device component.

Another issue, not explicitly addressed to any great extent in
treatments of interactions, is the
distinction between the two types of HMs. The study of the fully
polarized ferromagnet, also referred to as a saturated ferromagnet,
in which the minority spin states are empty, extends back to the seminal
work of Edwards and Hertz.\cite{EdwardsHertz} CrO$_2$ is the simplest
example of this: high spin S=1 Cr$^{4+}$ has no occupied minority
states\cite{CrO2}.
The ``gap'' in CrO$_2$ is several eV wide, between occupied O $2p$ states and
the unoccupied minority $3d$ bands. For low energy considerations,
dc transport for example, minority spin states are out of the picture.
The more common HM involves $d$ states on either
side of the gap as well as the continuum in the metallic
channel; the minority charge excitations are typically narrowly gapped;
in intermetallics a commonly ooccurring value is $\sim$0.5 eV and even in
oxides is rarely much greater.  In the
saturated case, there are simply no minority charge excitations to
consider.

There have been several studies based on model
Hamiltonians~\cite{review,Chioncel1,Nandini,HMKondo}. In particular,
work based on the $s$-$d$ exchange model predicted the existence of
nonquasiparticle (NQP) density of states near the Fermi level, which
(mathematically) arise from the branch cut of the self-energy arising from
electron-magnon interactions~\cite{review}. Chioncel {\it
et.al.}~illustrated the coincidence between NQP states and the imaginary
part of local spin-flip susceptibility in the framework of the
single-band Hubbard Hamiltonian and DMFT approach (see below) where
saturated ferromagnetism is stabilized by the additional magnetic spin
splitting mimicking the local Hund rule~\cite{Chioncel1}. An effective
spin Hamiltonian was derived to account for the temperature and disorder
dependence of the magnetic properties of half-metallic double
perovskites~\cite{Nandini}. Remarkably, Kondo screening was recently
shown to stabilize ferromagnetic order and further result in a half
metallic phase with minority-spin gap in the Kondo lattice model with
antiferromagnetic coupling~\cite{HMKondo}.

The LDA+DMFT (local density approximation plus dynamical mean-field
theory) approach that is becoming widely used to treat strong
interactions is not based on a model Hamiltonian, instead using LDA results as
the non-interacting system together with a self-interaction correction.
This approach has been applied by Chioncel
and coworkers\cite{Chioncel1,Chioncel2,Chioncel3} to evaluate the effect
of interactions on HMs.  One of the most well studied intermetallic HMs,
NiMnSb,\cite{NiMnSb} was the first application of this combined
technique.\cite{Chioncel1} The calculated spectral density contained
non-quasiparticle (NQP) states within the minority gap, but above, rather
than pinned to, the Fermi level, allowing it to survive as a HM.
However, further investigations of NiMnSb
and other Heusler alloys show that the magnetic moment per formula unit,
the NQP spectral weight, and the total DOS are insensitive or only weakly
sensitive\cite{Chioncel1} to correlation effects.
In striking contrast, correlation effects were found to play a vital
role in zincblende VAs, which is calculated to be a ferromagnetic semiconductor
({\it i.e.} gapped in {\it both} spin channels)
within LSDA or the generalized gradient approximation GGA, but
predicted to be a half-metal ferromagnet due to band shifts produced
by LSDA+DMFT.\cite{VAs} Strong
correlation effects are also obtained in magnetite, which is uncommon in
HMs in having a  majority-spin gap.\cite{magnetite}
In the full Heusler compound Mn$_2$VAl which is HM within LSDA,\cite{WehtWEP}
an LSDA+DMFT treatment of local interaction led to NQP states within
the gap but below the Fermi level,\cite{Chioncel3} which would not
degrade spintronics-related properties.
At present the impact of interactions
appears to be highly material specific. However, the models and the treatment of
the interactions (always approximate in some way) has varied widely,
so few questions are truly settled.

The DMFT approach, which treats on-site interactions
and dynamics in detail, has the shortcoming of neglecting intersite
correlations. Hence spinwaves, or even short range spin order, that
still contain strong correlations between neighbors, are replaced by
identical but uncorrelated spin fluctuations on each interacting site.
It is unclear to what degree the misrepresentation of these excitations
may affect the character of
the interacting spectral density.

In this paper we investigate the survival (or not) and character of HM
phases based on a bilayer Hubbard model with unequal interlayer hopping
for the two spin species. This model allows a substantial parameter
range in which the non-interacting density of states (DOS) has a gap in
only one spin channel, and will be described in detail in the next
section.  We note here that the layer index can be regarded equivalently as
an orbital or band index, so that the bilayer Hubbard Hamiltonian
provides a useful pedagogical link between single orbital and
multiorbital models.

In the case when the spin up and spin down band structures are identical
(i.e.~in the absence of underlying magnetic order), tuning of the interlayer
hybridization in such bilayer models has been demonstrated to drive a
variety of quantum phase transitions~\cite{bilayer}.  For example, at
half filling, the ground state can have antiferromagnetic long-range
order for small interlayer (interband) hopping $t_{\perp}$, and enter a
disordered valence bond phase with singlet correlations between
electrons on two layers, for large $t_{\perp}$.  Likewise, the system
can evolve through Mott insulating
transitions\cite{liebsch04,arita05,sentef09} as $t_{\perp}$ is altered.
In the doped system, there is a topological reconstruction of the Fermi
surface, which modifies the spin fluctuations and changes the
superconducting gap symmetry \cite{bulut92}.
Adopting spin-asymmetric interlayer hopping, our model introduces
new avenues of behavior to be illuminated.

\section{Model and methodology}
\subsection{Previous work}
Previous studies of HMs in the single-band Hubbard model
focused\cite{Chioncel1} mostly on the limiting case of saturated
ferromagnetism achieved through an external Zeeman magnetic field $B$.
With the underlying up and down spin bands being degenerate, the
effective spin chemical potentials $\mu_\sigma = \mu \pm B$ are chosen
to depopulate one of the species.  An alternate way to achieve a
half-metal, one in which the two spins species can still have the same
population so that the polarization is zero, is to alter the band
structure so that a gap opens for just one of the species (which we will
choose to be the ``down" spins).  For example, in a one band
tight-binding model on a bipartite lattice with band $\epsilon({\bf
k})$, one can incorporate an additional staggered potential
$V_{\mathbf{j}\sigma}=(-1)^{\mathbf{j}} \, V_\sigma$, where the
$(-1)^{\bf j}$ has opposite sign on the two sublattices.  This
alternating potential mixes momentum states ${\bf k}$ and ${\bf k +
\pi}$ and opens up a gap in the dispersion relation $E({\bf k}) = \pm
\sqrt{\epsilon({\bf k})^2 + V_\sigma^2}$.  By choosing $V_{\uparrow}=0$
and $V_\downarrow \neq 0$, for an appropriate choice of Fermi level, the
down spin species is insulating while the up species is metallic.

Such a staggered potential, however, couples strongly to
antiferromagnetism since it provides a one body energy which favors an
oscillating down spin density on the two sublattices.  The dominant
nature of the resulting magnetic response might obscure the
determination of the effect of $U$ on the transport properties.  For
example, for a half-filled square lattice Hubbard model which has a
divergent antiferromagnetic susceptibility $\chi_0(\pi,\pi)$, the additional staggered potential
immediately produces a state with long range antiferromagnetic order
(LRAFO) which is `trivial' in the sense that it does not arise from a
spontaneous breaking of symmetry, but rather from the externally imposed
potential.  This  LRAFO opens a Slater gap in the initially metallic up
spin spectrum, so that $U$ immediately, but in some sense trivially,
destroys HM behavior.

\subsection{Spin-asymmetric Hubbard model}
We avoid this confusing aspect by considering instead the slightly more
complex case of a two layer (or, equivalently, two orbital) square
lattice Hubbard model with spin-dependent inter-layer (inter-orbital)
hopping,
\begin{equation}
\begin{split}
\hat{H} = &-t \sum\limits_{\langle \mathbf{i}\mathbf{j} \rangle m
\sigma}
(c^{\dagger}_{\mathbf{i}m\sigma}c_{\mathbf{j}m\sigma}^{\vphantom{dagger}}+h.c.)
- \sum\limits_{\mathbf{i}m\sigma} (\mu-\sigma B) n_{\mathbf{i}m\sigma}
\\
&-
\sum\limits_{\mathbf{i}\sigma} t^{\vphantom{dagger}}_{\perp\sigma}
(c^{\dagger}_{\mathbf{i}1\sigma}c_{\mathbf{i}2\sigma}^{\vphantom{dagger}}+h.c.)
\\
&+ U \sum\limits_{\mathbf{i}m} (n_{\mathbf{i}m\uparrow}-\frac{1}{2})
(n_{\mathbf{i}m\downarrow}-\frac{1}{2}).
\end{split}
\label{H}
\end{equation}
Here the additional index $m=1,2$ labels two layers (orbitals) while
${\bf i,j}$ are site indices and $\sigma$ is the spin.  The first term
is an intralayer nearest-neighbor hopping.  We consider a square lattice
with intralayer hopping $t=1$ to set the energy scale.  $t_{\perp
\sigma}$ is a spin-dependent interlayer (interorbital) hybridization,
and $U$ is an on-site repulsion.  The terms coupling to the density are
a spin-independent chemical potential $\mu$ and a Zeeman field $B$.  The
repulsive on-site interaction term is written in particle-hole symmetric
(PHS) form so that at $\mu=B=0$ the system is half-filled (for each spin
species), even if $t_{\perp \uparrow} \neq t_{\perp \downarrow}$.

The noninteracting limit $U=0$ has two bands for each spin,
\begin{eqnarray}
\epsilon_{\sigma}^-(\mathbf{k}) &&= - t_{\perp\sigma}
- 2t(\cos k_{x}+\cos k_{y}).
\nonumber \\
\epsilon_{\sigma}^+(\mathbf{k}) &&= + t_{\perp\sigma}
- 2t(\cos k_{x}+\cos k_{y}).
\label{nonintbands}
\end{eqnarray}
For $t_{\perp\sigma} \leq 4t$ these two bands overlap, yielding metallic
behavior. However, for $t_{\perp\sigma} > 4t$, Eq.~\ref{nonintbands}
characterizes a band insulator with gap $2(t_{\perp \sigma} - 4t)$.
This Hamiltonian, and choice of $U=0$ band stucture, represents a
half-metal without polarization, and avoids the externally imposed
antiferromagnetism which would arise from a staggered potential.  The
price paid is the introduction of the extra layer (orbital) degree of
freedom $m$. The Hubbard Hamiltonian on a diamond chain
is a one dimensional analog in which the interplay of perpendicular hopping and $U$ can create a correlation-induced
half metal for certain fillings\cite{diamondchain}.

\subsection{Underlying magnetic order}
The manner in which magnetic order (a fundamental necessity for a HM)
is built into the model may affect behavior. In our model, magnetic
order is implicit in the spin-dependence of $t_{\perp\sigma}$ but is
otherwise unspecified. A signature aspect of our model is that for U=0
it is a specific realization of the schematic symmetric HMAF DOS,
presented for example in Fig.~1 of Ref.~[\onlinecite{RuddWEP}].
In their DMFT study of a fully polarized Bethe lattice Hubbard model,
Chioncel {\it et.al.} used a Zeeman field to mimic Hund's rule coupling,
thereby splitting the two spin directions until the minority band was
empty.\cite{Chioncel1} In LSDA+DMFT studies of suspected HM materials
(NiMnAs, FeMnSb, VAs), Chioncel and coworkers\cite{Chioncel1,Chioncel2,Chioncel3}
based their dynamical corrections on the spin-split LSDA bands.

The goal of this paper is to determine the effect of the electronic
correlation term $U$ in the model Hamiltonian Eq.~\ref{H}.
Specifically, we compute the spin-resolved spectral densities  as
functions of $t_{\perp\sigma},~U$, and temperature $T$.  We also study
the interplay of $U$ and interlayer hopping $t_{\perp\sigma}$ on the
antiferromagnetic correlations.  For simplicity we will set
$t_{\perp\uparrow}=0$ and vary $t_{\perp\downarrow}$ so that spin up
fermions are metallic and the spin down fermions can be tuned from metal
to band insulator at $t_{\perp\downarrow}=4t$.

At the PHS point $\mu=B=0$, both spin species are half-filled
(regardless of the values of $U,t,t_\perp$ or temperature $T$).  This
immediately implies the polarization is identically zero, so that our
model system realizes the exotic half-metallic antiferromagnetism
(HMAF)~\cite{HMAF}. Although there has been to date no clear
confirmation of novel HMAF materials, a variety of candidates have been
proposed, including La$_2$VCuO$_6$\cite{LVCO} as a likely member of the
double perovskite system with two magnetic ions,\cite{WEPhmaf}
semiconductors doped with dilute magnetic ions,\cite{diluted} and monolayer
superlattices CrS/FeS and VS/CoS\cite{Nakao}. An unusual semi-Dirac
half-semimetal arises in untrathin VO$_2$ films\cite{sD1,sD2}. Our
model's bands closely resemble the simplest realization of a HMAF as
arising from two exchange-split ions that are antialigned, shown
schematically and discussed in Ref. [\onlinecite{pickettreview}].
After considering the fundamental PHS case, we will also present some
results for non-zero Zeeman field $B$, where the system can be
expected to acquire
nonzero polarization.


\subsection{DQMC procedures}
We treat the interaction using the determinant quantum Monte Carlo
(DQMC) technique\cite{blankenbecler81,white89}.  DQMC is a numerically exact
approach to solve interacting tight binding electron Hamiltonians like
Eq.~\ref{H}.  In comparison with DMFT, DQMC has the advantage of being able
easily to incorporate and measure magnetic, charge, and pairing
correlations between different spatial sites.  On the other hand, DQMC
has the drawback of being formulated on finite spatial lattices so that
finite size effects must be assessed.
DQMC also is limited by the fermion ``sign problem''\cite{loh90}
(much more so than single site DMFT)
which prevents the acquisition of data at low temperatures.
Most of the results presented in this paper will be for two $8\times 8$
layers and inverse temperatures $\beta=6$.  We will show that the sign
problem is somewhat alleviated for Zeeman field $B\neq0$ so that
lower temperatures can be reached.

\subsection{Properties to be studied}
In order to distinguish metal from insulator, and see the effect of $U$
on the half-metallicity, we will examine the single-particle density of
states, $N_\sigma(\omega)$.  This quantity is obtained by an analytic
continuation of the local imaginary-time dependent Green's function
$G_{\sigma}(\tau)=\sum\limits_{{\bf j}} \langle T
c^{\phantom{\dagger}}_{{\bf j}\sigma}(\tau) c^{\dagger}_{{\bf
j}\sigma}(0) \rangle$, that is, by inverting,
\begin{equation}
   G_{\sigma}(\tau)= \int_{-\infty}^{\infty}\ d\omega \frac{e^{-\omega\tau}}{e^{-\beta\omega}+1}\ N_{\sigma}(\omega) \label{aw}
\end{equation}
using the maximum entropy method\cite{gubernatis91}.  Our focus will be
on the density of states at the Fermi surface $N_\sigma(\omega=0)$ to
see if HM behavior survives at non-zero $U$.

Although the system has no net polarizaton at $B=\mu=0$ from the
viewpoint of {\it total} up and down occupations, the distinction
between spin directions induces a polarization of the {\it conduction}
electrons, which is quantified by,
\begin{equation}
	P(E_{F}) = \frac{N_{\uparrow}(E_{F})-N_{\downarrow}(E_{F})}
{N_{\uparrow}(E_{F})+N_{\downarrow}(E_{F})} \,\,\,.
\end{equation}
This quantity has been the focus of much experimental work, since HMs
(at T=0) have 100\% polarization which is what makes them so attractive
for spintronics applications.

We also study the dc electrical conductivity $\sigma_{\rm dc}$, which is
extracted from the current-current correlation function,
\begin{equation}
   \Lambda_{xx}(\mathbf{k},\tau) = \sum_{\mathbf{i}}
e^{i\mathbf{k} \cdot \mathbf{l}}
\langle j_{x}(\mathbf{l},\tau)j_{x}(0,0) \rangle
\end{equation}
where $j_{x}(\mathbf{l},0) = it\sum_{\sigma}
(c^{\dagger}_{\mathbf{l}+x,\sigma}c_{\mathbf{l}\sigma}^{\vphantom{dagger}}
- c^{\dagger}_{\mathbf{l}\sigma}c_{\mathbf{l}+x,\sigma}^{\vphantom{dagger}})$.
The conductivity is obtained using the approximate form of the
fluctuation-dissipation theorem\cite{FD,GtauNw}, valid at large $\beta$,
\begin{equation}
 \Lambda_{xx}(\mathbf{k}=0,\tau=\beta/2) = \pi \sigma_{dc}/\beta^{2}
\end{equation}

The magnetic magnetic structure factor
\begin{equation}
  S(\mathbf{q})=\frac{1}{N} \sum\limits_{\bf l,j}
e^{i\mathbf{q}\cdot (\mathbf{l}-\mathbf{j})} \langle
(n_{{\bf l}\uparrow}-n_{{\bf l}\downarrow})
(n_{{\bf j}\uparrow}-n_{{\bf j}\downarrow}) \rangle
\end{equation}
is of interest as well.
In the ordered phase the spin correlations
$\langle
(n_{{\bf l}\uparrow}-n_{{\bf l}\downarrow})
(n_{{\bf j}\uparrow}-n_{{\bf j}\downarrow}) \rangle$
are non-zero for large ${\bf l-j}$ and $S({\bf q})$
grows linearly with
the lattice size $N$ at the appropriate ordering wave vector ${\bf q}$.

\section{Choice of parameter ranges}
Most HM magnets investigated to date are at most moderately correlated
intermetallic compounds\cite{review} and DFT treatments may be quite
realistic.  For instance, photoemission
experiments and resonant x-ray scattering has led to the estimate of
Hubbard interaction $U=2$ eV for NiMnSb~\cite{Uvalue}, while a Wannier
orbital analysis has shown that the bandwidth of the NiMnSb bands
crossing $E_{F}$ is $W\sim$ 4 eV.  The ratio $U/W \sim
1/2$ puts this HM compound in the weakly to moderately correlated
regime.  In our model system, the non-interacting bandwidth is $W =8t$ at
$t_\perp=0$, which suggests $U/t=0-4$ is the relevant range to study.

Correlation effects of this size can fundamentally change the physics of
related tight-binding Hamiltonians, {\it e.g.} opening a correlation gap or
inducing AFLRO and a Slater gap; see for example Ref. [\onlinecite{vollhardt12}]
and references therein.  The near-neighbor square lattice tight binding model is
unstable to arbitrarily small $U$, although this sensitivity is a
result of the van Hove singularity (vHs) and perfect nesting of that model.
The majority channel in our model however retains these same features
(at $\mu=0$ only, of course); in the minority channel the vHs are
shifted to either side of $\mu$=0. Even with one remaining vHs, we
will show below that interactions play a much less dramatic role
when one spin channel is gapped.

There is, after all, new physics in a HM as well as the
new phenomena discussed in Sec. I: the new energy scale
that is given by the spin-down gap $\Delta =
2(t_{\perp,\downarrow}-4t)$.  Down-spin charge excitations are gapped by
$\Delta$, which make interactions further differentiate the spins,
for example by enabling sharp `impurity' states (only) within the
spin-down gap. Spin-flip interactions are gapped by the separation
between $\mu$ (the up-spin Fermi level) and the nearest down-spin band
edge, which is $\Delta/2$ at $\mu$=0 in our model.  Kondo processes
vanish due to the absence of low energy spin-flip
excitations. Perhaps more relevantly, the vHs- and nesting-driven
spin density wave instability obviously will be quenched when there is
a minority gap.  We show it is also suppressed in the ferromagnetic
metal phase where the minority bands are split but not gapped.

\section{Effect of on-site interaction}

\begin{figure}
\psfig{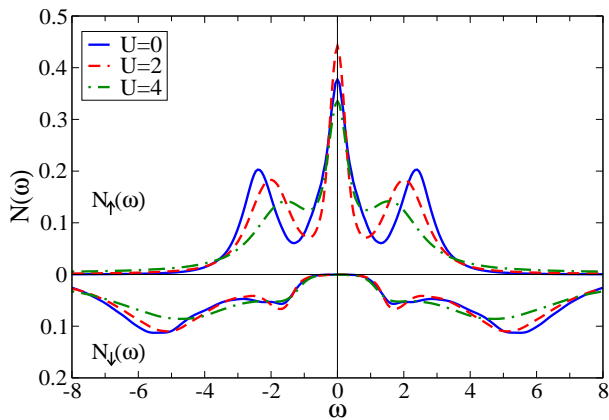}
\caption{Half metallic behavior persists in the presence of
interactions:  the down species density of states is virtually unaltered
at $U/t=4$ from its $U/t=0$ value.  Here $t_{\perp\uparrow}=0$ and
$t_{\perp\downarrow}=5$ so that the down species (only) has a
noninteracting band gap $\Delta=2t$.  The up species is metallic with a
density of states at the Fermi surface that is relatively insensitive to
$U$.  The lattice size $N=8\times8$ in each layer and the inverse
temperature $\beta=5$.  The chemical potential and
Zeeman fields $\mu=B=0$.
However, due to the particle-hole symmetry at half filling,
there is no signature associated with the NQP states.
}
\label{dosB5tup0tdn5}
\end{figure}

\begin{figure}
\psfig{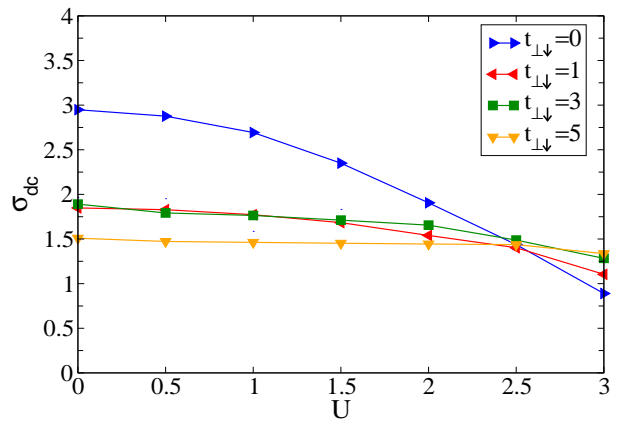}
\caption{
The dc conductivity for $\beta=6$ is shown as a function of $U$.
For the half metal
$t_{\bot\downarrow}=5t$, $U$ has very little effect on $\sigma_{\rm
dc}$, which is consistent with the invariance of the density of states
in Fig.~\ref{dosB5tup0tdn5}.  In a fully metallic phase $t_{\bot\downarrow} = 0$ there is a clear
decrease of $\sigma_{\rm dc}$ with $U$ due to the additional
electron-electron scattering which can occur when both species have
nonzero density of states at the Fermi surface.
}
\label{condB6}
\end{figure}

\begin{figure}
\psfig{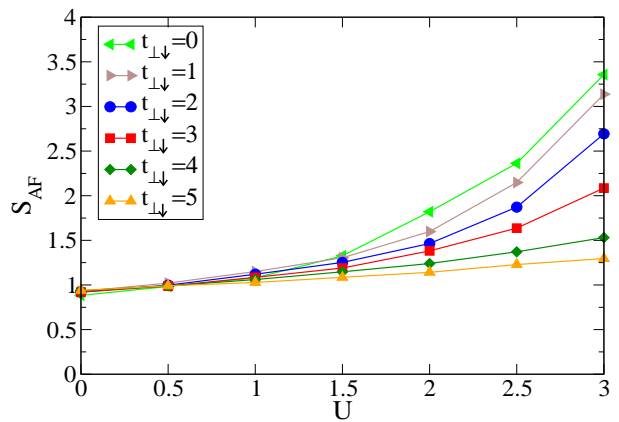}
\caption{Staggered static structure factor $S_{AF}$ through the
metal-HM transition. Increasing interlayer coupling reduces the
increase with increasing U until,
for half metallic phase $t_{\bot\downarrow}=5$, $U$ has little
effect on the (small) magnetic coupling.
In symmetric bilayers, large interlayer promote singlet formation,
which are suppressed here by the HM gap.
}
\label{safB6tup0}
\end{figure}

The moderating of strong interation effects is seen in the density of
states, shown in Fig.~\ref{dosB5tup0tdn5}, where the insulating species
$N(\omega)$ at $U/t=4$ is virtually indistinguishable from its $U=0$
form.  Here we have set $t_{\perp \uparrow}=0$ and $t_{\perp
\downarrow}=5t$ and $\mu=B=0$ so that the down spin species has a
noninteracting band gap $\Delta=2(t_{\perp\downarrow}-4t) = 2t$ and the
up species is metallic.  The gap magnitude and tailing of states into
the gap is unaltered up to U=4.  The up (metallic) species
$N_\uparrow(E_{\rm F})$ is also constant from $U/t=0$ to $U/t=4$ to
within the accuracy of the maximum entropy inversion of Eq.~\ref{aw}.
The peaks in $N_\uparrow(\omega)$ away from $\omega=0$ are affected by
the finite size effect of the $8\times 8$ lattice.

One distinguishing characteristic of interactions in HMs extensively
discussed in previous studies is the appearance of the non-quasiparticle
(NQP) states within the spectral gap -- the magnetic polaron effect.
These NQP states in previous studies have (1) appeared, at $\omega=0$,
above $\omega=0$, and below $\omega=0$, and (2) also possibly not appeared.
In the generic picture, the density of NQP states vanishes at the Fermi
level ($\omega=0$) but increases toward an energy scale of the order of
the magnon frequency, leading to an asymmetry of spectral function.
\cite{review} Assisted perhaps by the particle-hole
symmetry in our model (implying the symmetry of spectral function) at
half filling, NQP states may be inhibited from appearing, and there is no
signature of NQP states in Fig.~\ref{dosB5tup0tdn5}.  The implicit
nature of magnetic order in our model may also play a role, but other
treatments have also incorporated some implicit origin of the magnetic
order in a HM.

The HM character can be expected to become more or less evident
in a property specific manner, and we now describe a few examples.
The relatively minor effect of on-site
interaction $U$ is further evidenced by the behavior of the
dc-conductivity in Fig.~\ref{condB6}, where the interlayer coupling is
varied to move the system through the metallic phase
$t_{\perp\downarrow}$=2,~3, through the transition
$t_{\perp\downarrow}$=4, to the HM phase $t_{\perp\downarrow}$=5. At
$t_{\perp\downarrow}$=2 the conductivity is U-dependent because
$N(\omega\rightarrow 0)$ in the metallic phase has the underlying van
Hove singularity there. As U increases the $t_{\perp\downarrow}$
dependence weakens, and at $t_{\perp\downarrow}$=4 and 5, which is
crossing over into the HM phase, $\sigma_{dc}$ decreases and becomes
independent of the interaction strength: in the HM phase $t_{\perp
\downarrow}=5$, $\sigma_{\rm dc}$ falls by less than ten percent as $U$
increases from $U=1$ to $U=3$, and must be controlled solely by spin-up
processes.

Fig.~\ref{safB6tup0} demonstrates the correlation effects on the
staggered magnetic static structure factor, again for the progression
from metal to HM $t_{\perp\downarrow}$=2-5. The increase with U at small
interlayer coupling decreases as the HM phase is approached and entered,
and the variation of $S_{\rm AF}$ with U in the HM phase
$t_{\perp\downarrow}=5$ is quite small.  It is unlikely that
antiferromagnetic long-range order will arise at lower temperatures, and
it is known that, in the case of spin-independent interlayer
hybridization, $t_{\perp}$ drives a competing singlet ground state, with
a quantum critical point at  $t_{\perp} \sim 1.6$ above which the ground
state of the bilayer model no longer has AFLRO in its ground state.

In summary, our simple bilayer model of an unpolarized HM indicates the
robustness of the HM phase to interactions.  This is consistent with
some previous work, which concludes that the half-metallic properties of
several materials such as NiMnSb and other Heusler alloys are
insensitive or only weakly sensitive to correlation
effects~\cite{Chioncel1}.
It is not apparent whether the explicit particle-hole symmetry at half
filling and implicit nature of magnetic order in our model has any
substantial effect, but whatever the reason there is no signature
here associated with the states in the gap.

\section{Effect of temperature}

\begin{figure}
\psfig{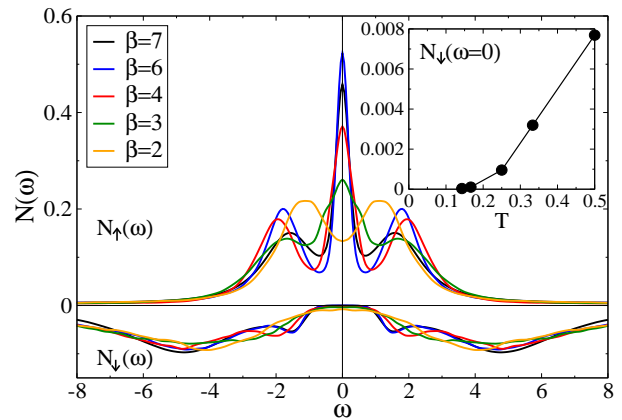}
\caption{Main panel:
Temperature effects on the density of states of half-metallic phase ($U=3$).
The inset shows $N_\downarrow(\omega=0)$ versus $T$.
The half-metal gap in $N_\downarrow(\omega)$ appears to be robust
for $\beta > 3 \,\,\, (T < t/3).$
}
\label{dostup0tdn5U3vsT}
\end{figure}

\begin{figure}
\psfig{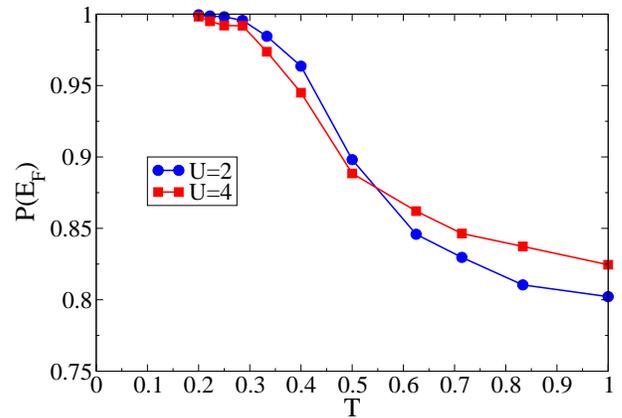}
\caption{The temperature dependence of the
conduction electron spin polarization, where the non-interacting
gap is equal to $\Delta$ = 2$t$ ($t$=1).
$P(E_{F})$
begins to turn downwards at $T \sim \Delta/6$, with minor dependence on
the value of $U$.
}
\label{PE}
\end{figure}

The temperature dependence of the half-metallic properties and their
stability against finite-temperature spin excitations are crucial for
practical applications. As mentioned before, one crucial effect is the
depolarization caused by finite-temperature.  With the
separate spin bands fixed in our model, the effect of reduced
magnetization upon approaching the Curie temperature is not included,
but we can study temperature effects ``at fixed magnetization.''

Fig.~\ref{dostup0tdn5U3vsT} illustrates the temperature effects on the
density of states in the half-metallic phase. While substantial
rearrangements of spectral weight occur for spin up, the primary effect
for spin down is an increased tailing of states across the band edges
and into the gap as T increases.  Only when $\beta<3$ is the
half-metallic feature destroyed, {\it i.e.}
$N_{\downarrow}(\omega\rightarrow 0)$ becomes appreciably non-zero.
This corresponds to a temperature $T \sim t/3 \sim \Delta /6$ since the
down spin band gap $\Delta = 2t$.  If we
phenomenologically introduce $\Delta(T)$ which vanishes at the Curie
temperature, and suppose that the magnitude of the gap is the primary
energy scale for this purpose, then we can infer destruction of HM
character by interactions (and thermal broadening) around $T^* =
\Delta(T^*)/6.$

We can also examine thermal effects by evaluating the conduction
electron spin polarization $P(E_{F})$.  The combined effects of
finite-temperature and Hubbard interaction $U$ are given in
Fig.~\ref{PE}.  The polarization begins to deviate downwards from unity
at $T \sim t/3$.
The role of $U$ on $P(E_F)$ is negligible in the low
temperature regime, but as $T$ increases to intermediate
temperatures $U$ has a distinct depolarization effect.
Thinking of our parameters as very roughly relevant
to NiMnSb, $t/3$ is a very high temperature, one at which the underlying
magnetic order (fixed in our model) physically has diminished drastically
or vanished. Even well above room temperature thermal effects
as well as effects due to interaction $U$ up to 4 on $P(E_F)$ are negligible.

\begin{figure}
\psfig{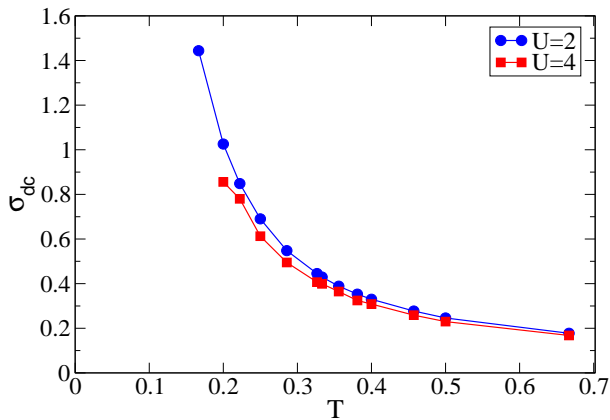}
\caption{Temperature dependence of conductivity, for U=2 and 4.
In the HM phase ($t_{\perp\downarrow}=5$) the conductivity (see Eq. 6)
is contributed only by the up spin
carriers, which as noted earlier is not affected strongly by these
values of U.}
\label{condvsT}
\end{figure}

A separate diagnostic of the nature of half-metallic phase and its
robustness against the Hubbard interaction $U$ is provided by the
conductivity, Fig.~\ref{condvsT}, evaluated from Eq. 6.  Recall that in
a half metal for these values of parameters, the conductivity arises only
from the majority (ungapped) channel, and spin-flip scattering is
frozen out at these temperatures leaving scattering only in the
charge channel. This data
supports the previous indications of the relatively small effects
of interactions in our bilayer model of a
half metal. The primary difference for $U=2$ is the significantly
larger conductivity when $T$ falls below $0.2t$, compared to
the $U=4$ trajectory. On the other hand,  $\sigma_{dc}$ at higher temperatures hardly depends on $U$.

\section{Effect of Zeeman magnetic field}

\begin{figure}
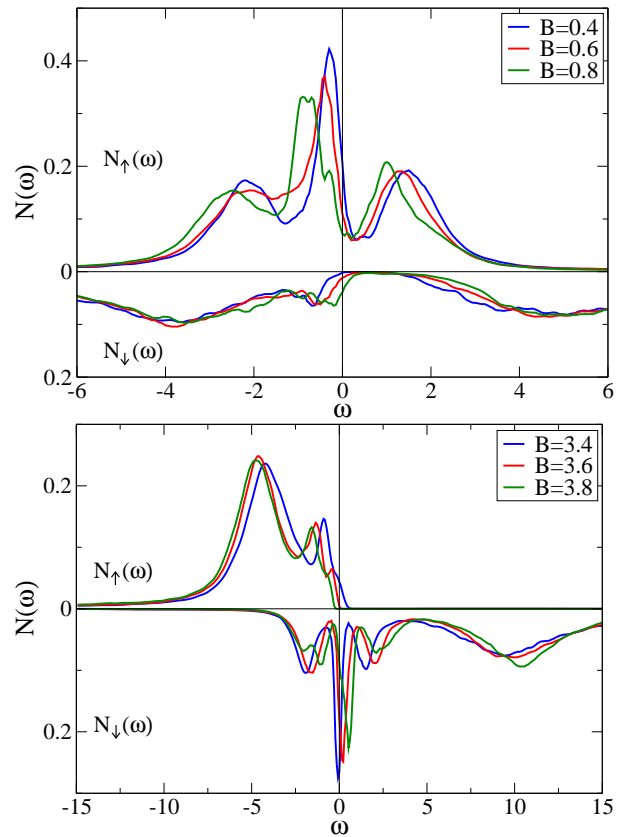

\psfig{figure=dosB8tdn5U3vsB1.eps,height=5.5cm,width=8.0cm,angle=0,clip}
\psfig{figure=dosB8tdn5U3vsB2.eps,height=5.5cm,width=8.0cm,angle=0,clip}
\caption{Effects of Zeeman field $B$ on the density of states, at
fixed $\mu$.  Here $U=3, \beta=8, t_{\perp \downarrow}=5$.
The Zeeman field $B$
shifts the spectral function of both spin species. \underbar{Top panel:}
At weak fields (viz. less than half the gap),
$N(\omega)$ more or less rigidly shifts with $B$ for
both spin species, as in mean field.
In the case of the insulating down spin electrons
this shift eliminates insulating behavior as $B$ approaches
half the band gap $2(t_{\perp}-4t)$ and the magnetization begins to change.
\underbar{Bottom panel:} At large magnetic field the up spin density of
states is driven completely below the Fermi level $\mu + B$ so that now
the up species is insulating.  Meanwhile the down spin density is
metallic because, with its larger bandwidth, the down spin Fermi level
$\mu - B$ is not yet completely below the down bands.  In this way one
has a HM in which the metal/insulator roles of the two spin species is
reversed.
 } \label{dosB6tdn5U3vsB} \end{figure}

All of results above are at $\mu=B=0$ which, by particle-hole
symmetry, guarantees half filling of both spin species:
$\langle n_{{\bf i}m\sigma}\rangle=0.5$.
A Zeeman field B introduces a spin bias, but one of the signatures
of a HM is that it is impervious to magnetic fields that are not too
large, that is, as long as the gap persists and $\mu$ does not cross a
gap edge.  This vanishing of the spin susceptibility is self-evident at the
mean field level, where B merely shifts the relative positions of the up
and down bands.   If the chemical potential (finally determined by state
filling in the metallic channel) does not cross either gap edge, there is
no reoccupation of states of either spin. In fact there is no change (up
to a constant) in the real-space potential for either spin, so the
(many-body) states themselves do not change. The energy changes only due
to the trivial magnetic energy term $-M B$ ($M$ is the net magnetization,
which is unchanging until $\mu$ crosses a band edge) which
is zero in our HMAF-type model.

The interest here is in interacting systems, and this anticipated
behavior of HMs has been supported by some rigorous results based on
ground state many-body wavefunctions and spin density functional
theory\cite{eschrig,vignale}. The theory provides for ranges of applied Zeeman
fields for which the ground state, and therefore the spin-density matrix
is unchanging, just as at the mean
field level. The values of the applied fields, positive and negative, at
which the magnetism changes thereby provides the gap edges, and their
difference provides the (interacting) fixed particle number gap.

In our calculations when $t_{\perp \uparrow} = t_{\perp \downarrow} $ and
$B=\mu=0$ there is no sign problem at any temperature.  This is a
consequence of a total correlation between the signs of the up and down
spin determinants, so that their product, the probability of the
configuration, is always positive.  Allowing
$\mu$ or $B$ to become nonzero induces a sign problem so that, normally,
simulations are much more challenging.  Here, however, with $t_{\perp
\uparrow} \neq t_{\perp \downarrow} $ the correlation between spin up
and spin down determinants has already been broken, which is why our
simulations in the earlier sections do not extend beyond $\beta \approx 6$.
We can access a similar range here when $B \neq 0$.

In mean-field level, the effect of the external Zeeman field
is only to shift the spectral functions of both spin species rigidly in opposite
directions.  It is therefore natural to expect a transition from
half-metallic phase to normal phase at $|B_c|$ equal to half the band gap. We
show that this transition survives in the presence of Hubbard
interaction $U$, but with spectral weight redistribution which is
not captured in MF in addition to renormalization of $B_c$.

Fig.~\ref{dosB6tdn5U3vsB} shows the effects of Zeeman field on the
spectral density. In the top panel the weak magnetic field-induced
shifting of the spectral weights of both spin species in the opposite
directions is clear, though the shifts begin to deviate from being rigid.
HM character survives to $B_c\approx$0.4, compared
to $B_c$=0.5 without interactions. This ``magnetic
field gap'' is renormalized by $\sim$ 20\% at U=3.
In the bottom panel, we demonstrate that a strong enough magnetic field (of order the
half the bandwidth, B $\sim$ $W/2$) can induce a situation in which the up spin electron
bands become completely filled (heuristically, the effective chemical
potential $\mu_{{\rm eff}\uparrow}=\mu+B$ lies
above the band), while the down spin electrons remain metallic:
$\mu_{{\rm eff}\downarrow}=-\mu+B$ still cuts across a region of nonzero
$N_\downarrow(\omega)$. Since the filled majority channel holds one electron, the
occupation of some minority states reflects the fact that the total density has
increased as the field is increased at fixed chemical potential $\mu$=0.
At fixed particle density, a large field will indeed produce a filled majority
band, in which the minority band must be empty (the non-zero spectral density
must lie only above $\mu$=0). This state is a saturated ferromagnetic insulator.
In this way, as $B$ increases, the sequence of
phases at fixed $\mu$ proceed from HM with majority carriers $\rightarrow$
metallic ferromagnet $\rightarrow$ HM with minority carriers. This sequene
converts, at
fixed density, to HM with majority carriers $\rightarrow$
metallic ferromagnet $\rightarrow$ saturated ferromagnet insulator.

\begin{figure}
\psfig{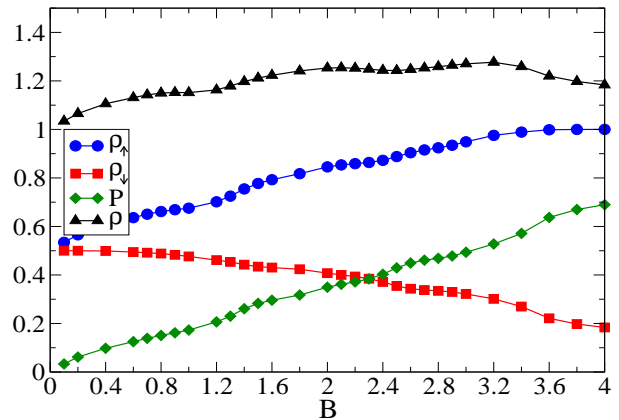}
\caption{Up, down, and total density as function
of Zeeman field at fixed $\mu=0$ and increasing $B$.  The system starts at
half-filling $\rho_{\uparrow} = \rho_{\downarrow}=0.5$ at $B=0$ and becomes
increasingly spin polarized as $B$ increases.  Ultimately at $B \sim 4$
equal  to half the band-width, the up bands are completely filled.  This
value of $B$ is not quite sufficient to empty the down band owing to its
slightly larger band-width. The polarization is $P=(\rho_{\uparrow}-\rho_{\uparrow})/(\rho_{\uparrow}+\rho_{\uparrow})$.
The parameters are chosen as $U=3, \beta=8$.
}
\label{magnetic}
\end{figure}

Another reflection of the magnetic behavior, again at fixed $\mu$=0,
is illustrated in Fig.~\ref{magnetic}.  The
external Zeeman magnetic field leads to an increasingly polarized
lattice [top panel, $P = (\rho_{\uparrow} - \rho_{\downarrow})/(\rho_{\uparrow} + \rho_{\downarrow})$],
as $\rho_{\uparrow}$ grows monotonically at the
expense of $\rho_{\downarrow}$.  At $B \sim 4$ the up bands are completely full
while the down bands, which due to their nonzero $t_{\perp \downarrow}$
have a larger bandwidth, are still crossed by the effective chemical
potential.  This appears to be re-entry into a (polarized) HM phase.
However, the spectra in Fig.~\ref{dosB6tdn5U3vsB} and the data of Fig.
\ref{magnetic} were obtained at constant $\mu$, and the B-induced increase
in density has be recognized but does not affect this picture.

\section{Summary}

We have used the numerically exact finite-temperature determinant
Quantum Monte Carlo (DQMC) method to study the
effect of strong interaction induced correlations on half-metallic
behavior in a multiband
Hubbard Hamiltonian.  Our model consists of a
bilayer square lattice, or, alternately viewed, a two orbital system,
with spin-dependent interlayer hybridization chosen to induce a band gap
in only one spin species.  This is a model appropriate to half metallic
antiferromagnets, since the lattice has the same number of spin up and
spin down electrons, but only the latter have a gapped non-interacting
spectrum.

By investigating the influence of an on-site Hubbard interaction $U$,
finite temperature $T$ and external Zeeman magnetic field $B$, we find
that the half-metallic properties are not particularly sensitive to the
interaction $U$, up to values equal to about half the bandwidth which
seem appropriate to intermetallic half metals.
Finite-temperature effects depolarize the conduction electron states only at
$T >t/3$ of the gap, with a degree of depolarization which depends weakly on
$U$.  A very large Zeeman magnetic field drives the system (at fixed
particle number) from half metal to metallic ferromagnet, and finally
to a ferromagnetic insulating phase when the minority spectral density
is Zeeman split completely above the majority spectrum.

An interesting issue in half-metallic ferromagnetic materials that remains undecided is
when non-quasiparticle (NQP) states arising from
electron-magnon interaction arise, and whether they are above, below, or
spanning the chemical potential.  The signature of such states is the
appearance of a resonance in the gapped channel.
In contrast to several earlier studies based on different models and using
a different treatment of the interaction, we have not seen any
distinct characteristic features associated with NQP states in the DQMC
studies of our model.  The difference may be related to the fact that our bilayer
system without the external magnetic field satisfies particle-hole
symmetry so that $E_{f}$ is always pinned to the center of the gap.  The
other possibility is that the sign problem has prevented us from
reaching low enough temperature to observe the development of the resonance, or
that the spin waves that may be required for a proper description of electron-magnon
interactions are not yet fully formed.
It is possible that the effects of interactions are simply model
and material dependent rather than universal.

\section{Acknowledgements}
We thank R.H.C. Peppers for useful input.
M.J. was supported by the National Science Foundation under
grant PHY-1005503, and R.T.S. and W.E.P. received support from
the Department of Energy Stewardship Science
Academic Alliances Program DE-NA0001842.


\end{document}